\documentclass[aps,prd,twocolumn,superscriptaddress,showpacs]{revtex4}

\usepackage{graphicx}
\usepackage{dcolumn}
\usepackage{bm}
\usepackage{epsfig}

\def \met {\mbox{${\not}{E_T}$} }
\def\bq{\begin{equation}}
\def\eq{\end{equation}}
\def\beq{\begin{eqnarray}}
\def\eeq{\end{eqnarray}}

\begin{document}

\title{First 
Measurement of the Ratio of Central-Electron to Forward-Electron  $W$ 
Partial Cross Sections in $p\bar{p}$
Collisions at $\sqrt{s}$ = 1.96 TeV.}

\affiliation{Institute of Physics, Academia Sinica, Taipei, Taiwan 11529, Republic of China} 
\affiliation{Argonne National Laboratory, Argonne, Illinois 60439} 
\affiliation{Institut de Fisica d'Altes Energies, Universitat Autonoma de Barcelona, E-08193, Bellaterra (Barcelona), Spain} 
\affiliation{Baylor University, Waco, Texas  76798} 
\affiliation{Istituto Nazionale di Fisica Nucleare, University of Bologna, I-40127 Bologna, Italy} 
\affiliation{Brandeis University, Waltham, Massachusetts 02254} 
\affiliation{University of California, Davis, Davis, California  95616} 
\affiliation{University of California, Los Angeles, Los Angeles, California  90024} 
\affiliation{University of California, San Diego, La Jolla, California  92093} 
\affiliation{University of California, Santa Barbara, Santa Barbara, California 93106} 
\affiliation{Instituto de Fisica de Cantabria, CSIC-University of Cantabria, 39005 Santander, Spain} 
\affiliation{Carnegie Mellon University, Pittsburgh, PA  15213} 
\affiliation{Enrico Fermi Institute, University of Chicago, Chicago, Illinois 60637} 
\affiliation{Comenius University, 842 48 Bratislava, Slovakia; Institute of Experimental Physics, 040 01 Kosice, Slovakia} 
\affiliation{Joint Institute for Nuclear Research, RU-141980 Dubna, Russia} 
\affiliation{Duke University, Durham, North Carolina  27708} 
\affiliation{Fermi National Accelerator Laboratory, Batavia, Illinois 60510} 
\affiliation{University of Florida, Gainesville, Florida  32611} 
\affiliation{Laboratori Nazionali di Frascati, Istituto Nazionale di Fisica Nucleare, I-00044 Frascati, Italy} 
\affiliation{University of Geneva, CH-1211 Geneva 4, Switzerland} 
\affiliation{Glasgow University, Glasgow G12 8QQ, United Kingdom} 
\affiliation{Harvard University, Cambridge, Massachusetts 02138} 
\affiliation{Division of High Energy Physics, Department of Physics, University of Helsinki and Helsinki Institute of Physics, FIN-00014, Helsinki, Finland} 
\affiliation{University of Illinois, Urbana, Illinois 61801} 
\affiliation{The Johns Hopkins University, Baltimore, Maryland 21218} 
\affiliation{Institut f\"{u}r Experimentelle Kernphysik, Universit\"{a}t Karlsruhe, 76128 Karlsruhe, Germany} 
\affiliation{High Energy Accelerator Research Organization (KEK), Tsukuba, Ibaraki 305, Japan} 
\affiliation{Center for High Energy Physics: Kyungpook National University, Taegu 702-701, Korea; Seoul National University, Seoul 151-742, Korea; SungKyunKwan University, Suwon 440-746, Korea} 
\affiliation{Ernest Orlando Lawrence Berkeley National Laboratory, Berkeley, California 94720} 
\affiliation{University of Liverpool, Liverpool L69 7ZE, United Kingdom} 
\affiliation{University College London, London WC1E 6BT, United Kingdom} 
\affiliation{Centro de Investigaciones Energeticas Medioambientales y Tecnologicas, E-28040 Madrid, Spain} 
\affiliation{Massachusetts Institute of Technology, Cambridge, Massachusetts  02139} 
\affiliation{Institute of Particle Physics: McGill University, Montr\'{e}al, Canada H3A~2T8; and University of Toronto, Toronto, Canada M5S~1A7} 
\affiliation{University of Michigan, Ann Arbor, Michigan 48109} 
\affiliation{Michigan State University, East Lansing, Michigan  48824} 
\affiliation{University of New Mexico, Albuquerque, New Mexico 87131} 
\affiliation{Northwestern University, Evanston, Illinois  60208} 
\affiliation{The Ohio State University, Columbus, Ohio  43210} 
\affiliation{Okayama University, Okayama 700-8530, Japan} 
\affiliation{Osaka City University, Osaka 588, Japan} 
\affiliation{University of Oxford, Oxford OX1 3RH, United Kingdom} 
\affiliation{University of Padova, Istituto Nazionale di Fisica Nucleare, Sezione di Padova-Trento, I-35131 Padova, Italy} 
\affiliation{LPNHE, Universite Pierre et Marie Curie/IN2P3-CNRS, UMR7585, Paris, F-75252 France} 
\affiliation{University of Pennsylvania, Philadelphia, Pennsylvania 19104} 
\affiliation{Istituto Nazionale di Fisica Nucleare Pisa, Universities of Pisa, Siena and Scuola Normale Superiore, I-56127 Pisa, Italy} 
\affiliation{University of Pittsburgh, Pittsburgh, Pennsylvania 15260} 
\affiliation{Purdue University, West Lafayette, Indiana 47907} 
\affiliation{University of Rochester, Rochester, New York 14627} 
\affiliation{The Rockefeller University, New York, New York 10021} 
\affiliation{Istituto Nazionale di Fisica Nucleare, Sezione di Roma 1, University of Rome ``La Sapienza," I-00185 Roma, Italy} 
\affiliation{Rutgers University, Piscataway, New Jersey 08855} 
\affiliation{Texas A\&M University, College Station, Texas 77843} 
\affiliation{Istituto Nazionale di Fisica Nucleare, University of Trieste/\ Udine, Italy} 
\affiliation{University of Tsukuba, Tsukuba, Ibaraki 305, Japan} 
\affiliation{Tufts University, Medford, Massachusetts 02155} 
\affiliation{Waseda University, Tokyo 169, Japan} 
\affiliation{Wayne State University, Detroit, Michigan  48201} 
\affiliation{University of Wisconsin, Madison, Wisconsin 53706} 
\affiliation{Yale University, New Haven, Connecticut 06520} 
\author{A.~Abulencia}
\affiliation{University of Illinois, Urbana, Illinois 61801}
\author{J.~Adelman}
\affiliation{Enrico Fermi Institute, University of Chicago, Chicago, Illinois 60637}
\author{T.~Affolder}
\affiliation{University of California, Santa Barbara, Santa Barbara, California 93106}
\author{T.~Akimoto}
\affiliation{University of Tsukuba, Tsukuba, Ibaraki 305, Japan}
\author{M.G.~Albrow}
\affiliation{Fermi National Accelerator Laboratory, Batavia, Illinois 60510}
\author{S.~Amerio}
\affiliation{University of Padova, Istituto Nazionale di Fisica Nucleare, Sezione di Padova-Trento, I-35131 Padova, Italy}
\author{D.~Amidei}
\affiliation{University of Michigan, Ann Arbor, Michigan 48109}
\author{A.~Anastassov}
\affiliation{Rutgers University, Piscataway, New Jersey 08855}
\author{K.~Anikeev}
\affiliation{Fermi National Accelerator Laboratory, Batavia, Illinois 60510}
\author{A.~Annovi}
\affiliation{Laboratori Nazionali di Frascati, Istituto Nazionale di Fisica Nucleare, I-00044 Frascati, Italy}
% ...
\collaboration{CDF Collaboration\footnote{With visitors from $^a$University of Athens, 
$^b$University of Bristol, 
$^c$University Libre de Bruxelles, 
$^d$Cornell University, 
$^e$University of Cyprus, 
$^f$University of Dublin, 
$^g$University of Edinburgh, 
$^h$University of Heidelberg, 
$^i$Universidad Iberoamericana, 
$^j$University of Manchester, 
$^k$Nagasaki Institute of Applied Science, 
$^l$University de Oviedo, 
$^m$University of London, Queen Mary College, 
$^n$University of California Santa Cruz, 
$^o$Texas Tech University, 
$^p$University of California Irvine, 
$^q$IFIC(CSIC-Universitat de Valencia), 
}}
\noaffiliation

\date{\today}

\begin{abstract}
\noindent 

We present a measurement of 
$\sigma(p\bar{p}\rightarrow{W}) \times
\mathcal{B}(W\rightarrow e \nu)$ 
at $\sqrt{s}$  = 1.96
TeV, using
electrons identified in the forward  region (1.2
$<|\eta|<$2.8) of the CDF II detector, in 223 pb$^{-1}$ of data. 
We measure  $\sigma\times \mathcal{B}$ = $2796\pm 
13$(stat)$^{+95}_{-90}$(syst)$\pm 162$(lum) pb.
Combining this result with a previous CDF measurement obtained using 
electrons in the central region ($|\eta| \lesssim 1$), 
 we present the first measurement of the ratio of  
central-electron to forward-electron  $W$ partial cross sections 
$R_{\mathrm{exp}} = 0.925\pm0.006$(stat)$\pm0.032$(syst), consistent with 
theoretical  predictions using 
CTEQ and MRST parton distribution 
functions.

\end{abstract}


\maketitle


The cross section for $W$ boson production in $p\bar{p}$ collisions has 
been computed 
at next-to-leading 
order 
(NLO)~\cite{nlo} 
and next-to-next-to-leading order (NNLO)~\cite{nnlo} in the 
strong coupling constant $\alpha_s$.
Experimental results can be used to  test the calculation of higher 
order QCD contributions and the parton distribution functions (PDFs) of 
the proton. 
The PDFs describe the momentum distributions of the elementary 
   constituents of the colliding hadrons, and are obtained from 
various parameterized  
fits to many sets of experimental data.
Their uncertainties 
   affect 
precision measurements like the masses and production 
cross
   sections of the $W$ boson and the top quark, at both the Tevatron and 
the 
   LHC~\cite{mlm}.  
Moreover, accurate 
   PDF modeling of the pseudorapidity $\eta$~\cite{eta} of 
the lepton from $W$ 
   decay is required for the use of $W$ production as a luminosity 
   monitor, an attractive option 
at the LHC~\cite{lummon}.
   
     The momentum fractions carried by the partons in colliding hadrons 
   determine the momentum distribution of the $W$ boson.  The $W$ boson 
 momentum  parallel to the proton beam direction cannot
   be measured in $p\bar{p}$ collisions, since the longitudinal momentum
   of the neutrino from the $W$ decay is not measured.  
A previous measurement of the forward-backward charge asymmetry of 
leptons 
from $W$s has been used to obtain some input on the momentum 
fraction dependence of the $u$ and $d$ quark PDFs 
within the proton~\cite{wasy}.
Independent 
   measurements of the $W$ cross section with central and forward leptons
   provide additional sensitivity to the $W$ rapidity $y_W$
   (Fig. 1), and are a novel way to constrain the PDFs.  We present the 
   first attempt to constrain PDFs using the ratio of $W$ boson cross 
sections 
   measured with central and forward electrons.  The largest experimental
   uncertainty, due to luminosity, cancels in this ratio.  We compare our
   measurement to the theoretical predictions obtained with two of the 
   most commonly used PDF sets.

\begin{figure}
 \begin{center}
\epsfig{file=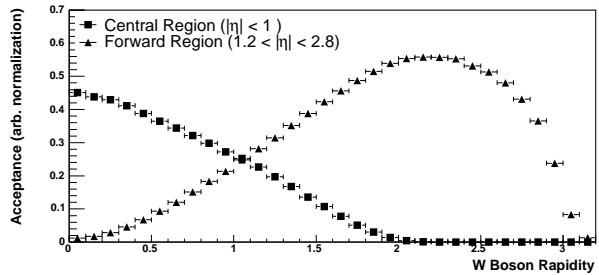,width=1.0\linewidth} 
\caption{Acceptance, obtained from simulation, as a function of the  $W$ 
boson rapidity, $A(y_W)$. ``Forward 
Region'' refers to this measurement (electron pseudorapidity 1.2 $<| \eta 
| <$ 2.8) while ``Central Region'' refers to the analysis reported 
in~\cite{wcdf} ($|\eta| <$1). 
The two analyses sample different regions of $y_W$. }
\label{w_acc}
\end{center}
\end{figure}

The $W$ cross section measurement presented in this Letter is obtained 
using 223 $\pm$ 13 pb$^{-1}$ of data  
collected by the CDF II detector
during Run II of the  Tevatron at $\sqrt{s}$ = 1.96 TeV.
$W$ bosons are identified by their
decays to electrons in the forward region (1.2 $< |\eta|<$ 2.8), from
which we obtain the inclusive cross section times branching 
fraction $\sigma({p}\bar {p}
\rightarrow W) \times \mathcal{B}(W\rightarrow e \nu)$.

Previous Run II results on $W$ production, based on electrons
with $|\eta | \lesssim$ 1, were reported by both the CDF 
and 
D\O~Collaborations~\cite{wcdf,wd0}.
In Run I, at $\sqrt{s}$ = 1.8 TeV,  D\O~reported a measurement based 
on 
electrons at $|\eta|<1.1$ and 
$1.5<|\eta|<2.5$~\cite{d0r1}, without 
separating the central from the forward regions.

The CDF II detector is described in detail 
elsewhere~\cite{detector}.
Tracking detectors inside  a 1.4 T solenoidal magnetic field 
reconstruct the charged particles' trajectories (tracks) and 
measure their 
momenta. The silicon tracking system (SVX)~\cite{silicon} 
provides precise 
measurement points from up to 8 radial layers of strip
sensors.
Outside the SVX is the central drift chamber
(COT), which
provides track measurements (hits) in 96
radial 
layers~\cite{cot}.
The COT allows full track reconstruction in the range $|\eta| <$ 
1.
The SVX 
extends the track reconstruction capability up to $|\eta| \simeq  2.8$.
Outside the tracking system, electromagnetic (EM) and hadronic 
(HAD)
 calorimeters  measure the energy of showering 
particles~\cite{calo}. 
In the forward region the position of the EM shower is measured 
by two layers of 
scintillating strips (PES)~\cite{pes}. 
The first layer of the forward EM 
calorimeter is used as a preshower detector~\cite{calo}.
Gas Cherenkov counters 
are used to determine the luminosity,
with 
an uncertainty of 5.8\%~\cite{clcn,lum}.
The trigger system has three  levels~\cite{l1,l3}. 
Data used in this analysis are selected by a 
trigger requiring missing transverse energy $\met >$ 
15 GeV 
and an EM cluster in the forward calorimeter with 
$E_T>$ 20 GeV. 

The offline selection of candidate $W$ decays begins 
by requiring
an energy 
cluster with $E_T >$ 20 GeV in the fiducial
region of the forward calorimeter at 1.2 $<|\eta|<$ 2.8.
The ratio of hadronic to electromagnetic energy deposition  must be 
 small: $E_{\mathrm{HAD}}/E_{\mathrm{EM}} < 0.05$. 
The EM cluster is required to be isolated~\cite{prd}.
The neutrino from the $W$ decay is identified by requiring $\met$ $>$ 25 
GeV.

To reduce the 
large 
remaining background, we compare 
the 
location and the energy deposition of the EM clusters in the calorimeter to
projections of  
three-dimensional tracks independently reconstructed by the tracking 
detectors.
Our track sample is dominated by tracks 
seeded 
by the SVX~\cite{ionim}. Typically, tracks in the region 1.2 $< |\eta| <$ 
1.6  have COT hit information, while those at larger $|\eta|$ do 
not. 
The candidate events are required
to have at least one track  
that extrapolates to the EM cluster shower 
centroid 
in the PES detector within 
3 cm in the $x$ and $y$ coordinates.
The selection is optimized by using a $Z\rightarrow ee$ data sample where 
one 
electron is detected in the central calorimeter~\cite{prd} and the other 
one in the 
forward calorimeter ($Z\rightarrow ee$ control sample). The 
probability for matching a track to an EM cluster 
detected in the forward region in $Z\rightarrow ee$ events is 
49.2 $\pm$ 0.5 \%. 
The $z$ coordinate of  the track 
intersection with the beam axis 
must be within  60 cm from the 
detector center. Finally, 
electron candidates must satisfy $E/p$ $<$ 2. 
After all requirements the sample contains 48,165 events. 

The kinematic and geometric acceptance ($A$) for $W\rightarrow e\nu$ 
events is determined using the {\sc pythia} event 
generator~\cite{pythia} and a full simulation of the CDF II detector 
based on 
the {\sc geant} 
simulation 
package~\cite{geant}.
We extract  
$A(y_W)$
from the simulation, and convolve it with a NNLO 
calculation of 
$d\sigma/dy_W$~\cite{dixon}.
We compute the 
central value of the acceptance using 
the MRST 2001 next-to-next-to-leading-log (NNLL) PDF set ~\cite{mrst} 
(in analogy with the $W$ cross section measurement in the central 
region~\cite{prd}) and find $A$ = 0.2567 $\pm$ 0.0002. 
 Two different sets of next-to-leading-log (NLL) PDFs are available 
(MRST01E and CTEQ6.1~\cite{cteq}).
 To encode uncertainties in the PDF data and  disagreements
between measurements included in the fits, the PDF sets provide  
eigenvectors formed in the space of the fit parameters (20 eigenvectors 
for CTEQ and 15 for MRST).   For each eigenvector, two complete PDF sets 
are provided corresponding to the changes in  each direction of that
eigenvector.
To estimate the uncertainty 
due to the choice of the PDFs we convolve $A(y_W)$ with the NLO
$d\sigma/dy_W$~\cite{prd} for each  PDF central value and $\pm 1\sigma$ 
eigenvalue. 
Using the CTEQ6.1 eigenvector basis set, we obtain a contribution to 
the acceptance uncertainty of ($+1.7$,$-1.3$)\%.
This value is 
roughly twice that obtained using the MRST01E PDF 
 set.
We use the difference between the NNLO  and NLO  $d\sigma/dy_W$ 
calculations to estimate the acceptance uncertainty due to higher-order 
QCD corrections ($\pm$ 0.47\%).
The other uncertainties on $A$  are described below.

The vector sum of the energy of hadrons recoiling against the 
$W$ boson enters the 
calculation of $\met$. 
We tune the detector response to these hadrons by applying scale factors 
and offsets to the components of their summed energy parallel 
 and perpendicular to the lepton momentum 
vector. We obtain a
systematic uncertainty of $\pm$ 0.35\% on $A$ by taking a 
variation corresponding 
to three standard deviations in the tuning parameters.

The energy scale and resolution modeling of electrons are calibrated with 
$Z\rightarrow ee$ events and result in an uncertainty of $\pm$ 0.24\%.  
The uncertainty on the scale as a function of $E_T$ is determined using 
the $E/p$ distribution.  The simulation models the $E_T$-dependence well, 
and we 
include a $\pm$ 0.26\% uncertainty on $A$ due to the 
statistical limitations of the constraint.  

 We vary the amount of material that an electron passes through by 
$\pm$1/3 
of a radiation length, 
based on measurements of electron 
energy deposition in the preshower detector. 
The resulting contribution to the acceptance uncertainty is $\pm$ 0.71\%.

       Differences in primary vertex reconstruction efficiency between 
data and simulation contribute less than 0.1\% to the acceptance 
uncertainty.  
 Finally, we 
vary the parameters of the {\sc pythia} model which influence the $W$ 
boson $p_T$ 
distribution~\cite{prd} within the constraints of a CDF Run~I $Z$ boson 
measurement,
and find the corresponding acceptance
uncertainty to be less than 0.1\%. 

Electron identification, track matching, and  $E/p$ 
efficiencies are measured 
using the $Z\rightarrow ee$ control sample.
The track matching 
efficiency is corrected
to account for the 
small kinematic differences of the $Z$ electrons with respect to those 
from $W$ decay
and the $\eta$ 
distribution of electrons coming from $W$s. 
These efficiencies
contribute the largest experimental	      
uncertainty to the cross section measurement, and are limited by 	      
the $Z$ statistics and the understanding of the background in 
the $Z\rightarrow ee$ sample.  The relative uncertainties on the cross section 
measurement from electron identification, track matching, and $E/p$ are 
2.0\%, 1.1\%, and 1.0\%, respectively.
The trigger efficiency is also measured from data, using independent 
triggers, and results in a relative uncertainty of 0.4\%. The overall 
efficiency is reported in Table I.

Backgrounds fall into two categories: multi-jet events, 
where one jet mimics an isolated high-$p_T$ electron and another jet is 
mismeasured in the calorimeters causing $\met$;
and electroweak backgrounds, $Z\rightarrow ee$ 
and $W\rightarrow \tau\nu$. 

The multi-jet background is estimated 
from data. Multi-jet 
events are 
characterized by significant energy in the cone around the electron and 
small $\met$~\cite{prd}.
We assume that these two variables are not correlated and  estimate the 
number of background events in the signal region  using control regions 
defined by either low $\met$ or  high energy in the isolation cone.  We 
vary the cuts on $\met$ and isolation  that define the control region and 
obtain a relative systematic uncertainty of 50\% on the multi-jet 
background estimate of 1.8\%.  We check this  calculation by determining 
the fraction of jets that pass our electron  criteria and applying this 
fraction to multi-jet events with large $\met$. We obtain good agreement.  

The electroweak backgrounds are estimated using simulation. 
We separately calculate the fraction of $Z\rightarrow ee$ 
and $W\rightarrow
\tau \nu$ events passing our selection. These fractions are 
then normalized to
data using the theoretical value for the ratio of $\sigma(Z)$/$\sigma 
(W)$~\cite{nnlo} and assuming 
lepton universality for the $W\rightarrow \tau \nu$ decays.
Background fractions from these processes are estimated to be 
2.2\% and 0.9\% for $W\rightarrow \tau\nu $ and $Z\rightarrow ee$ 
respectively. 

\begin{figure}[htb]
 \begin{center}
\epsfig{file=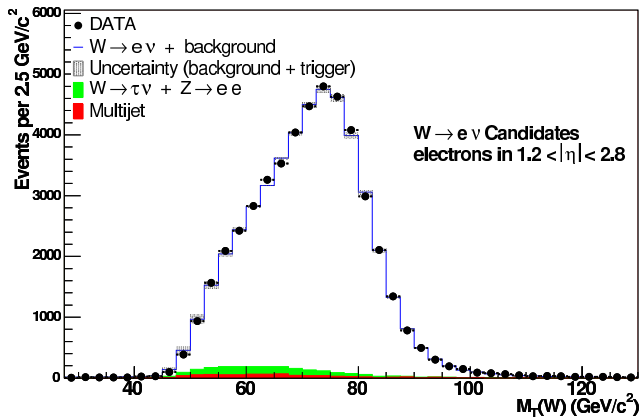,width=1.0\linewidth}
\caption{ $M_T$ distribution for  $W\rightarrow e\nu$
candidates (circles). The histogram is the sum of the expected background
and the predicted (from simulation) signal spectrum.}
\label{mtwfinal}
\end{center}
\end{figure}

We show the $M_T$~\cite{eta} distribution for both the signal 
and background 
contributions in Fig.~\ref{mtwfinal}.  The sum of signal simulation and 
background matches the data well.

From the number of selected events, 
the luminosity of the sample, the 
acceptance, efficiencies and 
backgrounds~\cite{prd} (see Table~\ref{tab:xsecef}),
\begin{table}[tb]
\begin{center}
\caption{ Geometric and 
kinematic acceptance, overall efficiency 
and 
expected number of background events.}
\label{tab:xsecef}
  \begin{tabular}{l l}
  \hline \hline \\ [-7pt]
  Background multi-jet & 846~~$\pm~57$~(stat)~~$\pm~423$~(syst)\\ [3pt]
  Background $Z\rightarrow ee$ & 417~~$\pm~5$~(stat) \\ [3pt]
  Background $W\rightarrow \tau \nu$ &1070~~$\pm~12$~(stat) \\ [3pt]
  \hline \\ [-7pt]
  Acceptance $A$ & 
0.2567~~$\pm~0.0002$~(stat)~~$^{+0.0051}_{-0.0042}$~(syst)\\ [4pt]
Efficiency $\epsilon_{\mathrm{TOT}}$ 
& 0.2863~~$\pm~0.0042$~(stat)~~$^{+0.0060}_{-0.0061}$~(syst) 
\\ [3pt] \hline \hline \\
\end{tabular}
\end{center}
\end{table}
we measure the inclusive cross section to be
$\sigma \times 
\mathcal{B} = 2796 \pm 13$(stat)$  ^{+95}_{-90}$(syst)$\pm 162$(lum) pb,  
consistent
with previous CDF results 
and with theoretical predictions~\cite{prd}.
The correct PDF must give the same total
$W$ cross section for central and forward electrons, within
statistical and systematic uncertainties.
It follows that the ratio 
of
partial cross sections  $\sigma_{p}$ = $\sigma \times 
\mathcal{B} \times 
A$, where $A$ is the kinematic and geometric acceptance, is equal to the 
true ratio of acceptances for the two regions. This experimental ratio can 
then be compared with acceptance ratios predicted by any set of PDFs.
The cross section based on central electrons, using the same PDF as used 
above for the 
forward electron measurement~\cite{prd},  is $\sigma \times \mathcal{B}$ = 
2771$\pm 14$(stat)$\pm 47$(syst) pb,
after removing uncertainties due to PDFs, luminosity, renormalization
scale and NLO/NNLO  effects.
The resulting  $\sigma_{p}$, measured for 
reconstructed electrons with 
$E_T >$ 25 GeV, $| \eta| \lesssim 1$~\cite{wcdf}
 and
\met $>$ 25 GeV is $\sigma^{\mathrm{cen}}_{p}$ = 
664$\pm 3$(stat)$\pm11$(syst) pb. In the forward 
region $\sigma_{p}$ for $E_T>$ 20
GeV, $ 1.2 <| \eta | < 2.8$, and \met $>$ 25 GeV is 
$\sigma^{\mathrm{for}}_{p}$ = 718$\pm
3$(stat)$\pm 21$(syst)~pb. 
All systematic uncertainties except those due to PDF and to NLO/NNLO 
effects, are assigned to $\sigma_{p}$.
Most of the luminosity uncertainty for the overlapping data-taking period 
cancels in 
the ratio, and we assign a 1\% 
systematic due to 
time-dependent luminosity uncertainty. 
All other uncertainties are uncorrelated. 
The experimental ratio is
 $R_{\mathrm{exp}}$ = $\sigma^{\mathrm{cen}}_{p}$ /   
$\sigma^{\mathrm{for}}_{p}$ = 0.925 $\pm 0.006$(stat)$\pm 0.032$(syst).
We compute also the
central-to-forward ratio of acceptances $R_{\mathrm{th}}$, obtained with 
two 
different PDF sets (CTEQ6.1 and MRST01E) at NLO level. 
For CTEQ6.1 the ratio is $R_{\mathrm{th}}$ = 
$0.924^{+0.023}_{-0.030}$(PDF) $\pm
0.004$(NLO/NNLO) and for MRST01E  
$R_{\mathrm{th}}$ = $0.941^{+0.010}_{-0.012}$(PDF) $\pm 0.004$(NLO/NNLO), 
where ``PDF'' indicates the uncertainty obtained by varying the 
eigenvalues relative to a given PDF set.

\begin{figure}[h]
 \begin{center}
\epsfig{file=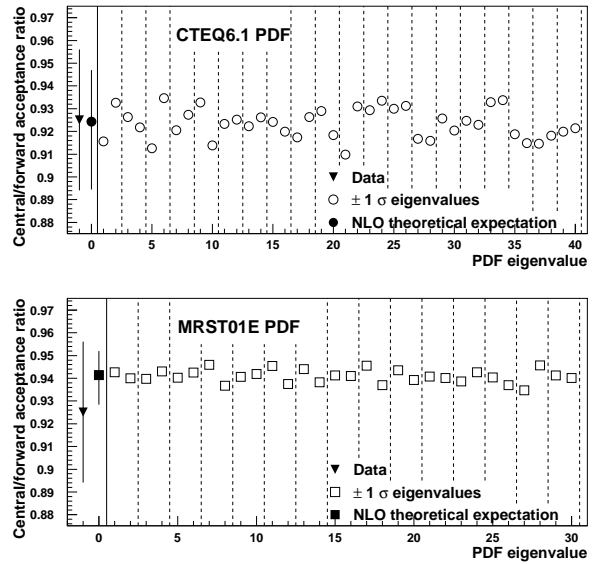,width=1.0\linewidth}
 \caption{ The ratio of central to forward 
acceptances, as a
function of the $\pm 1 \sigma$ eigenvalue of CTEQ6.1 (top) and MRST01E
(bottom) PDF sets. Dashed lines separate eigenvectors.}
\label{ratio}
\end{center}
\end{figure}

Figure~\ref{ratio} shows the experimental ratio of partial cross 
sections (solid triangles) compared 
to the CTEQ6.1 (upper plot) and MRST01E (lower plot) acceptance ratios 
(solid circle and square).
The data measurement is independent of PDFs.  
The ratios of acceptances are also computed varying each PDF 
eigenvalue by 
$\pm 1\sigma$ (giving 40  values for CTEQ6.1 and 30  
for MRST01E, shown as open circles and squares). 
Data and both PDF sets agree within  uncertainties, though the central 
values for MRST01E and CTEQ6.1 are slightly shifted with respect to each 
other.
The CTEQ6.1 has a larger uncertainty and some of the individual $\pm
1 \sigma$ eigenvalues show a sizeable deviation from the  central value.
 This is notably the case for eigenvector 1 (PDF eigenvalues 1 and 2 in 
Fig.~\ref{ratio}), in which the dominant contribution
is due to the $u$-valence quark, 
and
eigenvector 
3 (PDF eigenvalues 5 and 
6), in which
the most important contribution is due to the 
$d$-valence quark.
These eigenvectors 
impact $W$ boson 
measurements at the Tevatron, in particular the $W$ mass measurement.
Another large variation 
is visible for PDF eigenvector 5 (eigenvalues 9 and 10) 
in which the dominant contribution is due 
to sea quarks and gluons. This eigenvector is important for the $W$ 
rapidity distribution at the LHC.

Recently, a calculation of $R_{\mathrm{th}}$ at NNLO became 
available, which takes into account the spin correlation
between electron and neutrino, and the experimental selection of this
analysis. 
The authors find $R_{\mathrm{th}}=0.9266\pm0.0019$, in good
agreement with our measurement~\cite{meln}.

In summary, we have measured the $W$ inclusive production cross section 
with 
electrons identified at large pseudorapidities (1.2 $< |\eta| <$ 2.8) to 
be 
$\sigma \times \mathcal{B}$ = $2796\pm
13$(stat)$^{+95}_{-90}$(syst)$\pm 162$(lum) pb. We have measured a 
partial 
cross section using forward electrons 
$\sigma \times \mathcal{B} \times A$ = 
718 $\pm
3$(stat)$\pm 21$(syst) pb and the
ratio of central-electron to forward-electron partial cross sections  
$R_{\mathrm{exp}} = 0.925 \pm 0.006$(stat)$\pm  0.032$(syst). 
 
We thank the Fermilab staff and the technical staffs of the participating 
institutions for their vital contributions. This work was supported by the 
U.S. Department of Energy and National Science Foundation; the Italian 
Istituto Nazionale di Fisica Nucleare; the Ministry of Education, Culture, 
Sports, Science and Technology of Japan; the Natural Sciences and 
Engineering Research Council of Canada; the National Science Council of 
the Republic of China; the Swiss National Science Foundation; the A.P. 
Sloan Foundation; the Bundesministerium f\"ur Bildung und Forschung, 
Germany; the Korean Science and Engineering Foundation and the Korean 
Research Foundation; the Particle Physics and Astronomy Research Council 
and the Royal Society, UK; the Russian Foundation for Basic Research; the 
Comisi\'on Interministerial de Ciencia y Tecnolog\'{\i}a, Spain; the 
European Community's Human Potential Programme under contract 
HPRN-CT-2002-00292; and the Academy of Finland.


\begin{thebibliography} {*}
\bibitem{eta} CDF  uses two coordinate systems: cartesian and 
cylindrical. In the cartesian system $z$ is the beam axis, oriented in 
the proton direction, and $x$ and $y$ define the plane perpendicular to 
the $z$ 
axis. 
In the 
cylindrical system the $z$ axis is the same, $\theta$   
is the polar angle, $\phi$ is the 
azimuthal angle, and pseudorapidity $\eta = - \ln(\tan(\theta /2))$. 
Transverse energy and momentum are defined as $ E_T  = E \sin\theta$    
and $p_T = p \sin\theta$,   where $E$ is energy 
measured by the calorimeter and $p$ is momentum 
measured by the spectrometer. 
The missing $E_T$ ($\not\!\! \vec{E}_T$) is defined by
$\not\!\! \vec{E}_T$ = $- \sum_{i}  E_T^i \hat{n}_i$, where 
$\hat{n}_i$ is a 
unit   
vector perpendicular to the beam axis and pointing at the $ i^{th}$
calorimeter tower. We also define $\not\!\! E_T=|{\not\!\!\vec{E}_T}|$.
The transverse mass
is
: $M_T c^2 = \sqrt{2E_T\met (1 - \cos{\Delta\phi})}$,  
where $\Delta\phi$ is the azimuthal angle between the electron and the 
 $\met$.
\end{thebibliography}
\end{document}